\RequirePackage{fixltx2e}
\documentclass[aip,jcp,reprint,twocolumn,amsmath,amssymb,amsfonts,floatfix,letterpaper]{revtex4-1}
\usepackage{miniacs}

\setkeys{acs}{usetitle = true}
\usepackage{amsmath}
\usepackage{amssymb}
\usepackage{amsfonts}

\usepackage{mathptmx} 
\usepackage[scaled=0.90]{helvet} 
\usepackage{courier} 
\normalfont

\usepackage{graphicx}
\usepackage{braket}
\usepackage{color}
\usepackage{booktabs}
\usepackage{multirow}
\usepackage{array,contour} 

\newcommand{\emb}[1]{\mbox{$#1$}} 

\renewcommand{\d}{\mathrm{d}}  

\renewcommand{\vec}[1]{\mathbf{#1}}
\newcommand{\mat}[1]{\vec{#1}}

\renewcommand{\phi}{\varphi}
\renewcommand{\epsilon}{\varepsilon}

\newcommand{\norm}[1]{\left\Vert #1 \right\Vert}
\newcommand{\efrac}[2]{\emb{\frac{#1}{#2}}}
\newcommand{\vnabla}{\contour[2]{black}{$\nabla$}}

\renewcommand{\footnote}[1]{ [#1]}
\renewcommand{\onlinecite}[1]{\citenum{#1}}

\renewcommand{\phi}{\varphi}

\definecolor{HlColor}{rgb}{0.7,0.0,0.0}

\begin{document}

\title{\color{blue}{Efficient treatment of local meta-generalized gradient density functionals via auxiliary density expansion: the density fitting (DF) J+X approximation}\vspace*{0.18cm}}
\date{\today}
\author{Alyssa V. Bienvenu}
\author{Gerald Knizia}
\email{knizia@psu.edu}
\affiliation{Department of Chemistry, The Pennsylvania State University, University Park, PA 16802 (USA)}

\begin{abstract}
   We report an efficient technique to treat density functionals of the meta-generalized gradient approximation (mGGA) class in conjunction with density fitting of Coulomb terms (DF-J) and exchange-correlation terms (DF-X).
   While the kinetic energy density $\tau$ cannot be computed in the context of a DF-JX calculation, we show that the Laplacian of the density $\upsilon$ can be computed with almost no extra cost.
   With this technique, $\upsilon$-form mGGAs become only slightly more expensive (10\%--20\%) than GGAs in DF-JX treatment---and several times faster than regular $\tau$-based mGGA calculations with DF-J and regular treatment of the density functional.
   We investigate the translation of $\upsilon$-form mGGAs into $\tau$-form mGGAs by employing a kinetic energy functional, but find this insufficiently reliable at this moment.
   However, $\upsilon$ and $\tau$ are believed to carry essentially equivalent information beyond $\rho$ and $\norm{\vnabla\rho}$ [Phys.~Rev.~B {\bf 2007}, 75, 155109], so a reparametrization of accurate mGGAs from the $\tau$-form into the $\upsilon$-form should be possible.
   Once such functionals become available, we expect the presented technique to become a powerful tool in the computation of reaction paths, intermediates, and transition states of medium sized molecules.
\end{abstract}

\maketitle

\section{Introduction}

In DFT calculations of small and medium sized molecules, it is frequently possible to retain the most expensive electron repulsion integrals (ERIs) in main memory, particularly if the density fitting (DF) approximation is employed.\cite{whitten73,dunlap79,ffk1993,vahtras1993integral,eichkorn1995auxiliary,eichkorn1997auxiliary,weigend:rihf,weigend:RiVsCholesky}
In these cases, and particularly if efficient three-index ERI integration techniques are used,\cite{ahlrichs:3centerintegrals,kallay:3centerintegrals,werner:molpro} the total computational cost of a DFT calculation is strongly dominated by the terms associated with the exchange correlation contributions to the Fock matrix and the energy, which have to be numerically evaluated on a finite set of grid points.
One way to reduce this cost is to invoke the auxiliary density expansion (ADE), which was introduced by Laikov,\cite{Laikov:DensityExpansion} and has been picked up and extended by K\"oster and coworkers\cite{koester:AuxDensityXc2004,flores_moreno:AuxiliaryDensityPerturbationTheory,koster2009minmax,zuniga2016analytical} and others.\cite{birkenheuer2005model}
In this technique, rather than computing the density functional contributions from the exact density for the trial Kohn-Sham wave function $\ket{\Phi}$,
\begin{align}
   \rho(\vec r) = \sum_i^\mathrm{occ} n_i |\phi_i(\vec r)|^2 = \sum_{\mu\nu}^\mathrm{bf} \gamma^{\mu\nu} \chi_\mu(\vec r)\chi_\nu(\vec r),\label{eq:ExactDensity}
\end{align}
where the $\{\phi_i\}$ denote $\ket{\Phi}$'s occupied orbitals, $\{n_i\}$ their occupation numbers, $\{\chi_\mu\}$ the orbital basis functions, and $\{\gamma^{\mu\nu}\}$ the density matrix elements, in the ADE, the same auxiliary density $\tilde \rho(\vec r)$, as used to compute the Coulomb matrix $\mat j$  in the DF approximation, is also used to compute the density functional contributions.
Concretely, in the course of a DF-J computation, one determines the set of expansion coefficients $\{\gamma^F\}$ such that
\begin{align}
   \gamma^{F} &:= \sum_G [\mat J^{-1}]^{FG}  \sum_{\mu\nu} (G|\mu\nu)\gamma^{\mu\nu}\label{eq:gammaF}
\end{align}
where $F,G$ run over an auxiliary density fitting basis set, $[\mat J]_{FG} := (F|G)$ denotes the two-index Coulomb integrals, and $(F|\mu\nu)$ the regular three-index Coulomb integrals. \footnote{Here and in the following, multiplication with $\mat J^{-1}$ should be read as \emph{``solve an equation system $\mat J x = \mat b$ using a suitable matrix decomposition of $\mat J$''} (e.g., Cholesky or spectral decomposition); actual inverse matrices should \emph{not} be computed, as this negatively impacts numerical stability.\cite{knizia:numstab}}
One can show that this approximation leads to an approximate density
\begin{align}
   {\tilde \rho}(\vec r) = \sum_F \gamma^F \chi_F(\vec r),\label{eq:ApproxDensity}
\end{align}
which has the property that the self-interaction integral of the residual fitting error becomes minimal:
\begin{align}
   \frac{1}{2}(\tilde \rho - \rho|\tilde \rho - \rho) \rightarrow \mathrm{min},
\end{align}
which is equivalent to minimizing the square deviation of the electric field generated by the two densities $\rho$ (Eq.~\eqref{eq:ExactDensity}) and $\tilde\rho$ (Eq.~\eqref{eq:ApproxDensity}).\cite{m2003,tm2003}

While this density fitting approximation is only ``robust''\cite{m2003,dunlap:VariationalRobustDf,reine:VariationalRobustDf} with respect to the computation of Coulomb integrals, Laikov has shown\cite{Laikov:DensityExpansion} that, nevertheless, this approximation still allows for an efficient and reasonably accurate computation of density functional contributions, too.
And, somewhat unexpectedly, it is still entirely feasible to compute exact analytical gradients of the energy with respect to the nuclear positions.
At least in principle, this combination of attributes makes the ADE method highly appealing for fast DFT calculations with ``pure''  (i.e., non-hybrid) functionals on small and medium sized molecules.

Our understanding is that the main reasons preventing its large-scale deployment are two-fold:
First, we are not aware of any published large scale tests of the accuracy of this approximation (for both relative energies and geometries) in conjunction with standard basis sets, such as the def2-basis sets and associated fitting basis sets of Weigend.\cite{Weigend:def2SVP_def2TZVPP,Weigend:def2QZVPP,weigend:UniversalJfit,Weigend:UniversalJkfit}
Second, the nature of this approximation makes it impossible to compute the kinetic energy density $\tau$, 
\begin{align}
   \tau(\vec r) = \sum_i^\mathrm{occ} n_i \phi^*_i(\vec r) \left(-\efrac{1}{2}\Delta\right)  \phi_i(\vec r),
\end{align}
since this quantity depends on the individual occupied orbitals ($\{\phi_i\}$), and cannot be obtained from the density ($\rho$ or $\tilde\rho$) itself.
However, $\tau$ is used as input in the parametrization of almost every density functional of the meta-Generalized Gradient Approximation class (mGGA) in current use.

We here report two preliminary findings regarding these points: First, while the kinetic energy density $\tau$ cannot be computed in the ADE, the Laplacian of the density $\upsilon=\Delta \rho$, which formally carries the same quality of information as $\tau$ if combined with $\rho$ and $\norm{\vnabla \rho}$,\cite{perdew2007laplacian} not only can be calculated in the ADE, but for standard basis functions of contracted Gaussians multiplied by Solid Harmonics,\cite{helgaker:purplebook} its calculation can be made extremely efficient.
Second, we report preliminary findings regarding the accuracy of the ADE with standard basis sets, which suggest that the expansion error is small compared to the intrinsic accuracy of the DFT functionals, and that the method is particulary suitable for the calculation of geometries, while affording a 4--5 times speed up compared to regular mGGA calculations.

\section{Combined Density Fitting of $j$ and $xc$ Terms}

In the DF-JX (density fitting Coulomb + exchange-correlation) algorithm we describe here, we apply the regular Coulomb density fitting approximation\cite{m2003} to both the Coulomb term and the exchange-correlation terms\cite{Laikov:DensityExpansion} of both the energy and the Fock matrix contributions.
In this model, the Fock matrix and energy are given as
\begin{align}
   \mat f &= \mat h + \mat j + \mat v_\mathrm{xc}, \label{eq:FockMatrix}
\\ E &= \sum_{\mu\nu} \gamma^{\mu\nu} h_{\mu\nu} + E_\mathrm{nuc} + E_\mathrm{coul} + E_\mathrm{xc},\label{eq:KsTotalEnergy}
\end{align}
where $\mat h$ denotes the regular one-electron Core-Hamiltonian operator $\mat h = \mat t + \mat v_\mathrm{nuc}$, and $E_\mathrm{nuc}$ denotes the nuclear repulsion energy.
This expression differs from the exact Kohn-Sham expression by the fact that both the electron-electron Coulomb repulsion energy $E_\mathrm{coul}$ and the exchange-correlation energy $E_\mathrm{xc}$ are specified in terms of the (same) auxiliary density $\tilde\rho$ defined in Eq.~\eqref{eq:ApproxDensity}.
This leads to the following expressions for the remaining energy contributions in Eq.~\eqref{eq:KsTotalEnergy}:
\begin{align}
   E_\mathrm{coul} &= \frac{1}{2}\iint \frac{\tilde\rho(\vec r)\tilde\rho(\vec r')}{\norm{\vec r - \vec r'}}\, \d^3 r\,\d^3 r'
\\                 &= \frac{1}{2}\sum_{FG} \gamma^F (F|G) \gamma^G\label{eq:EcoulDfForm}
\\ E_\mathrm{xc} &= \int \epsilon\big(\tilde\rho(\vec r), \tilde\sigma(\vec r), \tilde\upsilon(\vec r)\big)\, \d^3 r.\label{eq:ExcDfForm}
\end{align}
In $E_\mathrm{xc}$, the density-derived intermediate quantities are
\begin{align}
   \tilde\sigma(\vec r) &:= [\vnabla {\tilde \rho}(\vec r)]\cdot[\vnabla {\tilde \rho}(\vec r)] \label{eq:AuxDensitySigma}
\\ \tilde\upsilon(\vec r) &:= \Delta {\tilde \rho}(\vec r).\label{eq:AuxDensityTau}
\end{align}
The components of $\mat j$ and $\mat v_\mathrm{xc}$ in Eq.~\eqref{eq:FockMatrix} are defined as energy derivatives with respect to the density matrix elements $\gamma^{\mu\nu}$:
\begin{align}
   [\mat j]_{\mu\nu} &:= \frac{\partial E_\mathrm{coul}}{\partial \gamma^{\mu\nu}} = \sum_F (\mu\nu|F) \gamma^{F}, \label{eq:DfCoulombMatrix}
\\ [\mat v_\mathrm{xc}]_{\mu\nu} &= \frac{\partial E_\mathrm{xc}}{\partial\gamma^{\mu\nu}}  = \sum_F (\mu\nu|F) \bigg(\sum_G [\mat J^{-1}]^{FG}\,v_G^\mathrm{xc} \bigg), \label{eq:DfExchangeMatrix}
\end{align}
where we defined the one-index exchange-correlation potential vector elements  $\{v_G^\mathrm{xc}\}$ via
\begin{align}
   v_G^\mathrm{xc} &:= \frac{\partial E_\mathrm{xc}}{\partial\gamma^G}. \label{eq:OneIndexVxc}
\end{align}
To obtain expressions \eqref{eq:DfCoulombMatrix} and \eqref{eq:DfExchangeMatrix}, we used that for quantities depending only on the one-index density vector $\{\gamma^F\}$ (which includes both $E_\mathrm{coul}$ and $E_\mathrm{xc}$ if given via Eqs.~\eqref{eq:EcoulDfForm} and \eqref{eq:ExcDfForm}), we can get the density matrix derivatives ${\partial}/{\partial \gamma^{\mu\nu}}$ via
\begin{align}
   \frac{\partial}{\partial \gamma^{\mu\nu}} &= \sum_F \frac{\partial \gamma^F}{\partial \gamma^{\mu\nu}} \cdot \frac{\partial}{\partial \gamma^{F}}
\\ &= \sum_F \bigg(\sum_G [\mat J^{-1}]^{FG} (G|\mu\nu)\bigg) \frac{\partial}{\partial \gamma^{F}}.
\end{align}
For the inner part of this expression, ${\partial \gamma^F}/{\partial \gamma^{\mu\nu}}$ is computed using Eq.~\eqref{eq:gammaF}.

Let us consider the computation of the exchange-correlation energy $E_\mathrm{xc}$ (Eq.~\eqref{eq:ExcDfForm}) and the components of 
its potential vector $\{v_G^\mathrm{xc}\}$ (Eq.~\eqref{eq:OneIndexVxc}).
We first approximate
\begin{align}
   E_\mathrm{xc} &= \int \epsilon\big(\tilde\rho(\vec r), \tilde\sigma(\vec r), \tilde\upsilon(\vec r)\big)\, \d^3 r
\\ &\approx \sum_g w_g \epsilon(\tilde\rho_g, \tilde\sigma_g, \tilde\upsilon_g),
\end{align}
where $\{(\vec r_g,w_g)\}$ denote the points and weights of a finite integration grid, and $\tilde\rho_g, \tilde\sigma_g, \tilde\upsilon_g$ denote the density intermediates evaluated at the respective grid points:
\begin{align}
   \tilde\rho_g &= \tilde\rho(\vec r_g) = \sum_F \gamma^F \chi_F(\vec r_g)\label{eq:RhoOnGrid}
\\  \vnabla \tilde\rho_g &= \vnabla\tilde\rho(\vec r_g) = \sum_F \gamma^F [\vnabla\chi_F(\vec r_g)]
\\ \tilde\sigma_g &= [\vnabla \tilde\rho_g]\cdot[\vnabla \tilde\rho_g]
\\ \tilde\upsilon_g &= \Delta \tilde\rho(\vec r_g) = \sum_F \gamma^F [\Delta \chi_F(\vec r_g)].\label{eq:UpsilonOnGrid}
\end{align}
(We define derivative operators to not act beyond square brackets).
Based on these, we get for $v_F^\mathrm{xc}$ (Eq.\eqref{eq:OneIndexVxc}):
\begin{align}
   v_F^\mathrm{xc}= \frac{\partial E_\mathrm{xc}}{\partial \gamma^F} &= \sum_g w_g \Big( \frac{\partial \epsilon}{\partial \rho} \chi_F(\vec r_g) + \frac{\partial \epsilon}{\partial \upsilon} [\Delta \chi_F(\vec r_g)]\notag
\\   &\qquad\qquad+ 2\frac{\partial \epsilon}{\partial \sigma}\big([\vnabla \tilde\rho_g]\cdot[\vnabla \chi_F(\vec r_g)]\big)\Big).\label{eq:ExPotentialVector}
\end{align}
The vector elements thus obtained are inserted into Eq.~\eqref{eq:DfExchangeMatrix} and used to compute the exchange-correlation matrix $\mat v_\mathrm{xc}$.
Note that, in practice, the Coulomb matrix $\mat j$ and the exchange-correlation potential matrix $\mat v_\mathrm{xc}$ should not be computed separately, but only their sum $\mat j+\mat v_\mathrm{xc}$---by combining
Eqs.~\eqref{eq:DfCoulombMatrix} and \eqref{eq:DfExchangeMatrix}, such that only one (expensive) contraction with the three-index fitting integrals $(\mu\nu|F)$ is required.\footnote{For this reason we defined the Coulomb energy via Eq.~\eqref{eq:EcoulDfForm}, rather than the more conventional expression $\sum_{\mu\nu} \gamma^{\mu\nu} j_{\mu\nu}$. The latter expression yields the same result as Eq.~\eqref{eq:EcoulDfForm}, but it requires the Coulomb matrix $\mat j$---and $\mat j$ will not be available if only $\mat j + \mat v_\mathrm{xc}$ is computed.}

\section{Efficient Evaluation of Basis Functions}
All of the contributions in Eqs.~\eqref{eq:RhoOnGrid} to \eqref{eq:ExPotentialVector} are significantly simpler than in the regular density expansion method, in which the exact density resulting from Eq.~\eqref{eq:ExactDensity} is employed for exchange-correlation terms.
As a consequence, when using DF-JX, the evaluation of the actual auxiliary basis functions $\chi_F(\vec r)$ and their derivatives in Eqs.~\eqref{eq:RhoOnGrid} to \eqref{eq:UpsilonOnGrid} on the grid points $\{\vec r_g\}$ becomes one of the computationally most expensive sub-steps of the entire DFT procedure.
For this reason, it is mandatory to process these contributions efficiently.

\begin{figure}
   \centering
   \rule{0.98\columnwidth}{2pt}
   \\[1ex]\includegraphics[width=.98\columnwidth]{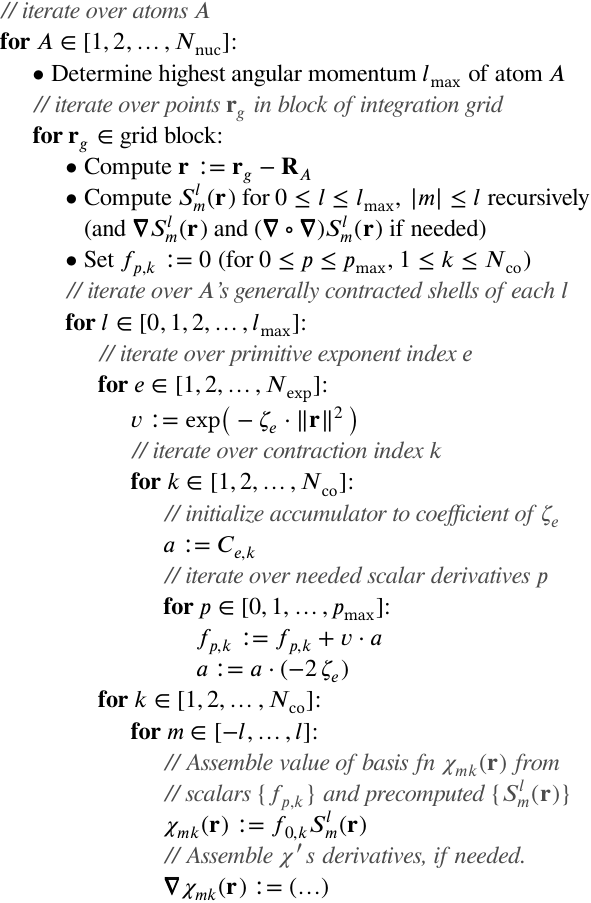}
   \rule{0.98\columnwidth}{2pt}
   \caption{Algorithm used to compute the values of basis functions $\{\chi_F(\vec r_g)\}$ over which the density $\tilde\rho(\vec r)$ in Eq.~\eqref{eq:ApproxDensity} is expanded on the points $\{\vec r_g\}$ of the integration grid.
   The quantities $\zeta_e$, $C_{e,k}$, and $f_{p,k}$ are defined before Eq.~\eqref{eq:ScalarFpk}.
   In the final step, the derivatives of $\chi$ are assembled using
   Eq.~\eqref{eq:NablaChi} and \eqref{eq:LaplaceChi} (for energy/Fock matrix evaluations, with $p_\mathrm{max}=2$) and Eq.~\eqref{eq:NablaNablaChi} and \eqref{eq:NablaLaplaceChi} (for analytic gradient evaluations, with $p_\mathrm{max}=3$). }
   \label{fig:BfOnGridAlgo}
\end{figure}

In our program we approach this as follows (see Fig.~\ref{fig:BfOnGridAlgo}):
First, we process all basis functions $F$ centered on the position in space $\vec R_A$ (typically an atom $A$) sequentially.
This allows us to only compute the contribution $\{S^l_m(\vec r_g - \vec R_A);\;l\in\{0,\ldots,l_\mathrm{max}\},\,m\in\{-l,\ldots,+l\}\}$ (and possibly its first, and in gradient computations, also second derivative) once for each atom, and employ the corresponding solid harmonic intermediates for all basis functions centered on the same atom.
We compute the $S^l_m(\vec r)$ using the standard recursive formulas.\cite{helgaker:purplebook}

Second, we iterate over each grid point $\vec r_g$, with its displacements $\vec r := \vec r_g - \vec R_A$ to the atom $A$.
For each generally contracted shell of basis functions on atom $A$,
defined by the primitive Gaussian exponents $\{\zeta_e;\,e\in\{1,\ldots,N_\mathrm{exp}\}\}$ and contraction coefficients $\{C_{e,k};\,e\in\{1,\ldots,N_\mathrm{exp}\}, k\in\{1,\ldots,N_\mathrm{co}\}\}$,
we compute the $p_\mathrm{max}\times N_\mathrm{co}$ matrix of scalar intermediates
\begin{align}
   f_{p,k} := \sum_{e=1}^{N_\mathrm{exp}} C_{e,k} (-2 \zeta_e)^p \exp(-\zeta_e r^2),\label{eq:ScalarFpk}
\end{align}
where $k\in\{1,\ldots,N_\mathrm{co}\}$ is the contraction index, and $p\in\{0,1,\ldots,p_\mathrm{max}\}$. $p_\mathrm{max}$ denotes the highest scalar function derivatives we require:
it is 1 for energy evaluations of GGAs, 2 for energy evaluations of mGGAs or gradients of GGAs, and 3 for gradients of mGGAs (note that only the $p=0$ term is computationally expensive---higher $p$ terms are obtained by simply multiplying the summands with $(-2\zeta_e)$).
With the $f_{p,k}$ thus defined, we obtain
\begin{align}
   \frac{\partial f_{p,k}}{\partial r_\alpha} = r_\alpha f_{(p+1),k},\qquad(r_\alpha\in\{x,y,z\})
\end{align}
or $\vnabla f_{p,k} = \vec r f_{(p+1),k},$
so Cartesian derivatives of these functions can be easily calculated.
This also allows computing their scalar Laplacian as
\begin{align}
   \Delta f_{p,k} &= \vnabla\cdot\left(\vnabla f_{p,k}\right) = \vnabla\cdot\left( \vec r f_{(p+1),k}\right) \notag
\\      &= \vec r\cdot\underbrace{[\vnabla f_{(p+1),k}]}_{\vec r f_{(p+2),k}} +  \underbrace{[\vnabla\cdot\vec r]}_{3} f_{(p+1),k} \notag
\\      &=  r^2 f_{(p+2),k} + 3 f_{(p+1),k}. 
\end{align}

From these $\{f_{p,k}\}$ and $S^l_m(\vec r)$ intermediates, the values of the basis functions $\chi_{mk}(\vec r)$ and their derivatives can be assembled as follows
(we omit the indices $l,A$ from $\chi_{mk}$ to simplify notation).
First, note that the value of $\chi(\vec r)$ itself is
\begin{align}
   \chi_{mk}(\vec r)&=S_{m}^{l}(\vec r) f_k(\vec r^2) = f_{0,k} S_{m}^{l}(\vec r),
\end{align}
and that, for the derivatives, we obtain
\begin{align}
  \vec{\vnabla} \chi_{mk}(\vec r)&=[\vec{\vnabla} f_{0,k}] S_{m}^{l}(\vec r) + f_{0,k} [\vec{\vnabla}S_{m}^{l}(\vec r)]
\\\Delta \chi_{mk}(\vec r)&= [\Delta f_{0,k}] S_{m}^{l}(\vec r) +  2[\vec{\vnabla} f_{0,k}]\cdot[\vec{\vnabla}S_{m}^{l}(\vec r)] \notag
\\&\quad+  f_{0,k} [\Delta S_{m}^{l}(\vec r)]
\\  &= [\Delta f_{0,k}] S_{m}^{l}(\vec r) + 2[\vec{\vnabla} f_{0,k}]\cdot[\vec{\vnabla}S_{m}^{l}(\vec r)].\label{eq:MysteriousVanishingOfLaplaceSlm}
\end{align}
For the second-to-last line, we used the fact that the solid harmonics $S^l_m(\vec r)$ fulfill the homogeneous Poisson equation,\cite{weniger:TheSphericalTensorGradientOperator} meaning that their Laplacian vanishes.
Inserting the previous results for gradient and Laplacian of $f_{p,k}$, this yields
\begin{align}
  \vec{\vnabla} \chi_{mk}(\vec r)&= f_{1,k} \vec r\, S_{m}^{l}(\vec r) + f_{0,k} [\vec{\vnabla}S_{m}^{l}(\vec r)]\label{eq:NablaChi}
\\\Delta \chi_{mk}(\vec r)&= 2f_{1,k}\,\vec r\cdot [\vec{\vnabla}S_{m}^{l}(\vec r)]  + \left(r^2 f_{2,k} + 3 f_{1,k} \right)S_{m}^{l}(\vec r).\label{eq:LaplaceChi}
\end{align}

\section{Analytic Gradients of the Energy}
For computing gradients of the energy, also the second derivatives of $\chi(\vec r)$ are required (just as in the regular GGA case), and the first derivatives of their Laplacian $\Delta\chi(\vec r)$:
\begin{align}
   (\vnabla\circ\vnabla) \chi_{mk}(\vec r) &= f_{0,k} [(\vnabla\circ\vnabla)S_{m}^{l}(\vec r)] \notag
\\       &\quad+ f_{1,k}\left( [\vnabla S^l_m(\vec r)]\circ\vec r + \vec r \circ [\vnabla S^l_m(\vec r)] \right)\notag
\\       &\quad +  \left(f_{1,k} \mat {\hat i}  + f_{2,k} (\vec r \circ \vec r) \right) S_{m}^{l}(\vec r)\label{eq:NablaNablaChi}
\end{align}
where $\circ$ denotes the outer product (dyadic product) between two vectors, and $\mat {\hat i}$ is the unit dyad (from $\vnabla\circ\vec r=\mat {\hat i}$); and
\begin{align}
   \vnabla \Delta \chi_{mk}(\vec r) &= 2 f_{2,k} \vec r (\vec r \cdot \vnabla S^l_m(\vec r)) + 2 f_{1,k} (\vec r \cdot \vnabla)(\vnabla S^l_m(\vec r))  \notag
\\ &\quad+\vec r \left(5 f_{2,k} + r^2 f_{3,k}\right) S^l_m(\vec r) \notag
\\ &\quad+ \left(5 f_{1,k} + r^2 f_{2,k}\right) [\vnabla S^l_m(\vec r)].\label{eq:NablaLaplaceChi}
\end{align}
Using these, the $E_\textrm{df-jx} = E_\mathrm{xc} + E_\mathrm{coul}$ contributions to the nuclear gradient of the energy can be written as\cite{Laikov:DensityExpansion}
\begin{align}
   E^{q}_\textrm{df-jx} &= E^{(q)}_\mathrm{xc} + \sum_{\mu\nu F} \gamma^{\mu\nu} (\mu\nu|F)^{(q)} \left(\gamma^F + d^F_\mathrm{xc}\right)\notag
\\   &\qquad - \frac{1}{2} \sum_{FG}\gamma^F (F|G)^{(q)} \left(\gamma^G + 2\,d^G_\mathrm{xc}\right)
\\ d^F_\mathrm{xc} &= \sum_G [\mat J^{-1}]^{FG} v_G^\mathrm{xc}
\end{align}
with $v_G^\mathrm{xc}$ from Eq.~\eqref{eq:ExPotentialVector} and $\gamma^G$ from Eq.~\eqref{eq:gammaF}.
As usual, in these expressions $q$ runs over the $3\,N_\mathrm{nuc}$ Cartesian components of the nuclear positions $\{\vec R_A,\,A\in\{1,\ldots,N_\mathrm{nuc}\}\}$, the $(\cdot)^{(q)}$-superscript denotes partial derivatives with respect to $R_q$, and $(F|G)^{(q)}$ and $(\mu\nu|F)^{(q)}$ denote derivative integrals.
The contribution from the grid integration of $E_\mathrm{xc}$ is given by
\begin{align}
   E^{(q)}_\mathrm{xc} &= \sum_g w_g \bigg(
   \frac{\partial \epsilon}{\partial \tilde\rho} \frac{\partial \tilde\rho(\vec r_g)}{\partial R_{q}} + 
   \frac{\partial \epsilon}{\partial \tilde\sigma}\frac{\partial \tilde\sigma(\vec r_g)}{\partial R_{q}} +
   \frac{\partial \epsilon}{\partial \tilde\upsilon}\frac{\partial \tilde\upsilon(\vec r_g)}{\partial R_{q}}\bigg)\label{eq:GridIntegrationDeriv}
\end{align}
where
\begin{align}
   \frac{\partial \tilde\sigma(\vec r_g)}{\partial R_{q}} &= \frac{\partial [\vnabla{\tilde\rho(\vec r_g)}]\cdot[\vnabla{\tilde\rho(\vec r_g)}]}{\partial R_{q}} 
\\ &= 2\,[\vnabla{\tilde\rho(\vec r_g)}]\cdot \frac{\partial [\vnabla{\tilde\rho(\vec r_g)}]}{\partial R_{q}} \label{eq:SigmaDeriv}
\\ \vnabla{\tilde\rho(\vec r_g)} &= \sum_F \gamma^F [\vnabla \chi(\vec r_g)].
\end{align}
In Eqs.~\eqref{eq:GridIntegrationDeriv} and \eqref{eq:SigmaDeriv}, the partial derivatives with respect to the nuclear coordinate $R_{q}$ are given via expressions such as
\begin{align}
   \frac{\partial \tilde\rho(\vec r_g)}{\partial R_{q}} = \sum_F \gamma^F \frac{\partial \chi_F(\vec r_g)}{\partial R_{q}}.
\end{align}
While, formally, this involves a sum over all fitting basis functions $F$, in practice the basis function derivative ${\partial \chi_F(\vec r_g)}/({\partial R_{q}})$ can only be non-zero if the nuclear coordinate $(q)$ refers to is identical with the nucleus on which the basis function $\chi_F$ is centered.
Therefore, for each $F$, there are only three nuclear derivatives $(q)$ to which it contributes (namely, ${\partial}/({\partial [\vec R_{A}]_x})$, ${\partial}/({\partial [\vec R_{A}]_y})$, and ${\partial}/({\partial [\vec R_{A}]_z})$, where $A$ is the atom index on which basis function $F$ is placed).
The derivatives of the basis functions themselves, which enter into these expressions, are given by Eqs.~\eqref{eq:NablaNablaChi} and \eqref{eq:NablaLaplaceChi}.
We here neglect contributions from the nuclear derivatives of the grid weights $\{w_g\}$;\cite{baker:NoGridWeightDerivatives} if required, these can be handled as usual.
As usual, we also assumed self-consistence of the wave function; in this case the total Kohn-Sham energy Eq.~\eqref{eq:KsTotalEnergy} is stationary with respect to the $\{\gamma^F\}$ and $\{\gamma^{\mu\nu}\}$ matrix elements, and their nuclear derivatives can be omitted.\cite{Laikov:DensityExpansion}

\section{Numerical Tests of Efficiency and Accuracy}
In this particular way of evaluating $\upsilon$-based mGGA functionals, no higher than first derivatives of the basis functions need to either be computed or assembled in order to compute the energy and Fock matrix contributions, and no higher than second derivatives are needed for the gradient of the energy.
These requirements are the same for GGA functionals, and this is only made possible by the realization that the Laplacian of the solid harmonics vanishes (Eq.~\eqref{eq:MysteriousVanishingOfLaplaceSlm}).
In combination, this makes the use of $\upsilon$-based mGGAs only slightly more expensive ($\approx$10\%--20\%) than the use of GGAs in the DF-JX formalism, and several times faster than regular $\tau$-based mGGA calculations (see Tab.~\ref{tab:timing} for timings).

\begin{figure}[t]
   \includegraphics[scale=0.27]{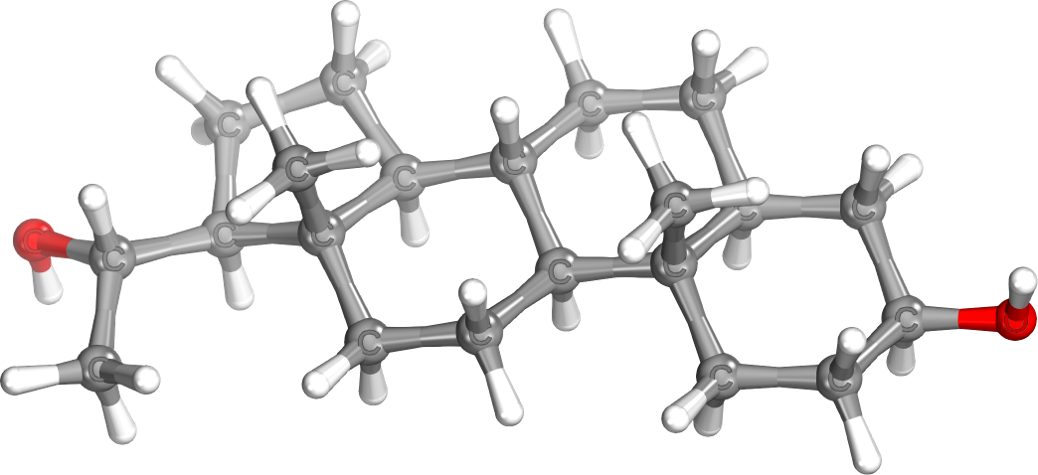}
   \caption{Pregnanediol: a medium sized organic molecule (59 atoms) used for timing comparisons (Tab.~\ref{tab:timing}). Number of basis functions: 929 (def2-TZVP\cite{Weigend:def2SVP_def2TZVPP}), 1523 (univ-JFIT\cite{weigend:UniversalJfit}); a reasonably sized integration grid with 229338 grid points was used.}
   \label{pic:preg}
\end{figure}

\begin{table}[t]
   \centering
   \caption{Wall-time comparisons for PBE (with auxiliary density expansion/DF-JX), TPSS (with regular density expansion/DF-J), and LL-TPSS (with DF-JX) for pregnanediol, with the orbital and fitting basis sets described in Fig.~\ref{pic:preg}. ``Wave function (total)'' includes time for integrals (3.3\,s), grid construction (0.6\,s), initial guess, and 10 SCF iterations.
   Computations were done with MicroScf v20171026 on a 2012 Lenovo W530 notebook with a 4-core Intel i7-3820QM CPU (2.70\,GHz), running Fedora 26 Linux, and employed OpenMP shared-memory parallelization over 8 threads.}
   \label{tab:timing}
   \vspace*{1ex}
   \begin{tabular}{c@{\hspace*{1em}}c@{\hspace*{1em}}c@{\hspace*{1em}}c} 
   \toprule
   \textbf{Timing\,/\,[s]} & \textbf{PBE} & \textbf{TPSS} & \textbf{LL-TPSS}\\
   \midrule
   \multicolumn{1}{l}{One SCF iteration} & \phantom{0}1.91 & \phantom{0}11.15 & \phantom{0}2.20 \\ 
   \multicolumn{1}{l}{Wave Function (Total)\hspace*{1em}} & 25.15 & 127.36 & 28.38 \\ 
   \multicolumn{1}{l}{Analytic Gradient} & \phantom{0}6.17 & \phantom{00}8.46 & \phantom{0}7.26 \\ 
   \bottomrule
   \end{tabular}
\end{table}

\begin{table}
   \caption{Test set of reaction energies and their RHF/AV6Z and CCSD(T)/CBS[56] reference values [kcal/mol]; taken from Ref.~\onlinecite{knizia:CcsdF12}}
   {
      \vspace*{.5ex}
      \begin{tabular}{lrr}
      \toprule
                                                      \multicolumn{1}{c}{\textbf{Reaction}\phantom{$T^{X}$}}&\multicolumn{1}{c}{\textbf{CCSD(T)}}&  \multicolumn{1}{c}{\textbf{RHF}}\\
      \midrule
      \makebox[1.4em][l]{1}                               {CO + H$_2$ $\rightarrow$ HCHO} &      -5.17\phantom{$^a$}&        0.25
  \\  \makebox[1.4em][l]{2}                    {CO + H$_2$O $\rightarrow$ CO$_2$ + H$_2$} &      -6.42\phantom{$^a$}&        0.06
  \\  \makebox[1.4em][l]{3}              {CH$_3$OH + HCl $\rightarrow$ CH$_3$Cl + H$_2$O} &      -8.03\phantom{$^a$}&      -6.00
  \\  \makebox[1.4em][l]{4}                             {H$_2$O + CO $\rightarrow$ HCOOH} &      -9.02\phantom{$^a$}&       -1.81
  \\  \makebox[1.4em][l]{5}           {CH$_3$OH + H$_2$S $\rightarrow$ CH$_3$SH + H$_2$O} &      -10.90\phantom{$^a$}&      -7.69
  \\  \makebox[1.4em][l]{6}           {CS$_2$ + 2 H$_2$O $\rightarrow$ CO$_2$ + 2 H$_2$S} &      -11.32\phantom{$^a$}&     -29.23
  \\  \makebox[1.4em][l]{7}                   {C$_2$H$_6$ + H$_2$ $\rightarrow$ 2 CH$_4$} &      -18.09\phantom{$^a$}&      -21.22
  \\  \makebox[1.4em][l]{8}                 {HNCO + H$_2$O $\rightarrow$ CO$_2$ + NH$_3$} &      -20.54\phantom{$^a$}&      -23.15
  \\  \makebox[1.4em][l]{9}                {CH$_4$ + Cl$_2$ $\rightarrow$ CH$_3$Cl + HCl} &      -23.46\phantom{$^a$}&     -26.29
  \\  \makebox[1.4em][l]{10}                        {Cl$_2$ + F$_2$ $\rightarrow$ 2 ClF } &     -27.07\phantom{$^a$}&     -34.04
  \\  \makebox[1.4em][l]{11}                         {CO + Cl$_2$ $\rightarrow$ COCl$_2$} &     -27.44\phantom{$^a$}&      -13.51
  \\  \makebox[1.4em][l]{12}           {CO$_2$ + 3 H$_2$ $\rightarrow$ CH$_3$OH + H$_2$O} &     -28.17\phantom{$^a$}&     -28.37
  \\  \makebox[1.4em][l]{13}                        {HCHO + H$_2$ $\rightarrow$ CH$_3$OH} &     -29.42\phantom{$^a$}&     -28.56
  \\  \makebox[1.4em][l]{14}                        {CO + 2 H$_2$ $\rightarrow$ CH$_3$OH} &     -34.59\phantom{$^a$}&     -28.31
  \\  \makebox[1.4em][l]{15}                {C$_2$H$_4$ + H$_2$ $\rightarrow$ C$_2$H$_6$} &     -39.55\phantom{$^a$}&     -39.16
  \\  \makebox[1.4em][l]{16}                  {SO$_3$ + CO $\rightarrow$ SO$_2$ + CO$_2$} &     -43.25\phantom{$^a$}&     -38.11
  \\  \makebox[1.4em][l]{17}                         {H$_2$ + Cl$_2$ $\rightarrow$ 2 HCl} &     -45.55\phantom{$^a$}&     -50.99
  \\  \makebox[1.4em][l]{18}                {C$_2$H$_2$ + H$_2$ $\rightarrow$ C$_2$H$_4$} &     -49.32\phantom{$^a$}&     -51.72
  \\  \makebox[1.4em][l]{19}          {SO$_2$ + H$_2$O$_2$ $\rightarrow$ SO$_3$ + H$_2$O} &     -50.42\phantom{$^a$}&     -55.38
  \\  \makebox[1.4em][l]{20}                 {CO + 3 H$_2$ $\rightarrow$ CH$_4$ + H$_2$O} &     -64.71\phantom{$^a$}&     -59.01
  \\  \makebox[1.4em][l]{21}                {HCN + 3 H$_2$ $\rightarrow$ CH$_4$ + NH$_3$} &     -76.73\phantom{$^a$}&     -80.16
  \\  \makebox[1.4em][l]{22}                  {H$_2$O$_2$ + H$_2$ $\rightarrow$ 2 H$_2$O} &     -87.25\phantom{$^a$}&     -93.54
  \\  \makebox[1.4em][l]{23}              {CO + H$_2$O$_2$ $\rightarrow$ CO$_2$ + H$_2$O} &     -93.67\phantom{$^a$}&     -93.48
  \\  \makebox[1.4em][l]{24}            {2 NH$_3$ + 3 Cl$_2$ $\rightarrow$ N$_2$ + 6 HCl} &     -97.38\phantom{$^a$}&     -115.38
  \\  \makebox[1.4em][l]{25}                {3 N$_2$H$_4$ $\rightarrow$ 4 NH$_3$ + N$_2$} &     -104.94\phantom{$^a$}&     -112.49
  \\  \makebox[1.4em][l]{26}                           {H$_2$ + F$_2$ $\rightarrow$ 2 HF} &     -135.03\phantom{$^a$}&     -145.97
  \\  \makebox[1.4em][l]{27}      {CH$_4$ + 4 H$_2$O$_2$ $\rightarrow$ CO$_2$ + 6 H$_2$O} &    -290.70\phantom{$^a$}&    -315.09
  \\  \makebox[1.4em][l]{28}              {2 NH$_3$ + 3 F$_2$ $\rightarrow$ N$_2$ + 6 HF} &    -365.82\phantom{$^a$}&    -400.32
  \\
      \bottomrule
      \end{tabular}
   }
   \label{tab:TestReactions}
\end{table}

\begin{table}[b]
 \centering
 \caption{Accuracy comparisons of density expansions.
   RMSD denotes the root mean square deviation between Method~1 and 2 for the 28 reactions listed in Tab.~\ref{tab:TestReactions}, MAD the mean absolute deviation, and MAX the maximum absolute deviation.
   DFT calculations are performed with def2-QZVPP\cite{Weigend:def2QZVPP} orbital basis sets and accurate integration grids. univ-JFIT and univ-JKFIT denote Weigend's universal Coulomb fitting basis sets\cite{weigend:UniversalJfit} and Coulomb/Exchange fitting basis sets\cite{Weigend:UniversalJkfit}.
   All energies given in $\mathrm{kcal}/\mathrm{mol}$.}
 \label{tab:ClosedShellReactions}
 \vspace*{.5ex}
 \begin{tabular}{l@{\hspace*{1em}}lrrr}
 \toprule
 \multicolumn{1}{l}{\textbf{Method 1}} & \multicolumn{1}{l}{\textbf{Method 2}} & \multicolumn{1}{l}{\hspace*{0em}\textbf{RMSD}} & \multicolumn{1}{l}{\hspace*{0em}\textbf{MAD}} & \multicolumn{1}{l}{\hspace*{0em}\textbf{MAX}} 
 \\ \midrule
               \multicolumn{5}{c}{\textit{Base accuracy of DF-JX for GGAs}}
      \\[.3ex] PBE (DF-J/JFIT)                         &PBE (DF-JX/JFIT)                        &      0.56&      0.41&      1.83
      \\       PBE (DF-J/JKFIT)                        &PBE (DF-JX/JKFIT)                       &      0.33&      0.24&      0.82
      \\[.8ex] \multicolumn{5}{c}{\textit{Deviation of TPSS and LL-TPSS (univ-JFIT)}}
      \\[.3ex] LL-TPSS (DF-J)                          &LL-TPSS (DF-JX)                         &      1.16&      0.73&      4.23
      \\       LL-TPSS (DF-J)                          &TPSS (DF-J)                             &      4.11&      3.23&      9.28
      \\       LL-TPSS (DF-JX)                         &TPSS (DF-J)                             &      4.42&      3.59&      9.46
      \\[.8ex] \multicolumn{5}{c}{\textit{Deviation of TPSS and LL-TPSS (univ-JKFIT)}}
      \\[.3ex] LL-TPSS (DF-J)                          &LL-TPSS (DF-JX)                         &      0.82&      0.62&      2.13
      \\       LL-TPSS (DF-J)                          &TPSS (DF-J)                             &      4.09&      3.21&      9.21
      \\       TPSS (DF-J/JFIT)                        &TPSS (DF-J/JKFIT)                       &      0.06&      0.04&      0.22
      \\[.8ex] \multicolumn{5}{c}{\textit{Deviation from high-level reference (univ-JFIT)}}
      \\[.3ex] TPSS (DF-J)                             &CCSD(T)/CBS                             &     12.06&      7.81&     42.83
      \\       LL-TPSS (DF-J)                          &CCSD(T)/CBS                             &     13.69&      9.12&     46.59
      \\       LL-TPSS (DF-JX)                         &CCSD(T)/CBS                             &     14.07&      9.19&     49.33
      \\[.8ex] PBE (DF-J)                              &CCSD(T)/CBS                             &      9.97&      6.87&     33.90
      \\       PBE (DF-JX)                             &CCSD(T)/CBS                             &      9.80&      6.73&     33.83
      \\       PBE (DF-J)                              &TPSS (DF-J)                             &      5.46&      4.46&     10.84
   \\ \bottomrule
   \\[-1ex] \multicolumn{5}{l}{{\small DF-J $\Rightarrow$ regular density expansion with DF for $\mat j$ and $E_\mathrm{coul}$}}
   \\ \multicolumn{5}{l}{{\small DF-JX $\Rightarrow$ auxiliary density expansion (Eq.~\eqref{eq:ApproxDensity})}}
   \\ \multicolumn{5}{l}{{\small LL $\Rightarrow$ Laplacian-level functional (translated via PC07)}}
\end{tabular}
\end{table}

The described methods have been implemented in the MicroScf program, which is also used in the following test calculations. MicroScf is used as integrated DFT driver of IboView\cite{knizia:IboViewHp,knizia:CurlyArrows} and will be released as a stand-alone open-source program in due time.
In the course of this work, the PC07 kinetic energy functional,\cite{perdew2007laplacian} the TPSS $\tau$-mGGA functional,\cite{tao2003climbing} and the LL-TPSS $\upsilon$-mGGA functional (\emph{vide infra}), as well as MN15-L (a non-separable $\tau$-mGGA functional),\cite{truhlar:mn15l} have been implemented with the help of the Maxima computer algebra system\cite{maxima} in order to derive the required density functional derivatives and translate them into efficient C++ code.
We here focus on PBE (GGA) and TPSS ($\tau$-mGGA) as representatives of their respective functional classes.
[While also other first-principles mGGA functionals were recently developed,\cite{sun:SCANmGGA,tao:AnotherMgga2016}
the results of novel approaches to functional development\cite{margraf:CorrelatedSingleParticleTheories,bartlett:PowerOfExactConditions2017,jin:TheQtpFamilyOfFunctionals2016}
as well as the increasing knowledge about the nature and limits of DFT functionals themselves\cite{cohen:FractionallyChargedNucleiVsElectronicStructure,cohen:ExactFunctionalTwoSideHubbard,mori:ComputingTheExactFunctional}
and impacts of their technical realization\cite{swart:SensitivityOfReactionsToFunctionalsAndTech} indicate a need for further empirical tests on relevant chemical systems before strong conclusions regarding their preferability over TPSS can be drawn.]
The test set of chemical reactions has been taken from Ref.~\onlinecite{knizia:CcsdF12} (including their RMP2/AVTZ geometries), and includes all closed-shell reactions for which CCSD(T) reference values at the extrapolated\cite{helgaker:l3extrapolation1,helgaker:l3extrapolation2} AV5Z/AV6Z\cite{Dunning:ccpVnZ_HBCOFNe,Kendall:AVnZ_HBCOF,Dunning:VnZ_plus_d} basis set limit were available.
The concrete reactions and their reference values are listed in Tab.~\ref{tab:TestReactions}.

In order to test the accuracy of the described methods,
various methods combinations were compared to each other, as reported in Tab.~\ref{tab:ClosedShellReactions}.
In these calculations, we employed Weigend's (accurate) def2-QZVPP orbital basis sets\cite{Weigend:def2QZVPP} and large DFT integration grids, in order to probe the intrinsic accuracy of the reported approaches.
We first test the auxiliary density expansion (Eq.~\eqref{eq:ApproxDensity}) itself (in the following denoted as DF-JX), compared to the regular density expansion (Eq.~\eqref{eq:ExactDensity}) in combination with the standard density fitting approximation applied only to Coulomb terms (this combination is in the following denoted as DF-J).
Comparison between PBE (DF-JX) and PBE (DF-J) on one side, and of LL-TPSS (DF-JX) and LL-TPSS (DF-J) on the other side shows that the auxiliary density expansion itself is reasonably accurate, with RMSDs of the reaction energies below 1.2 kcal/mol in all tested cases.
These deviations are, in particular, entirely negligible when compared to the intrinsic (in)accuracies of the density functional results themselves with reference to high-level wave function results from CCSD(T) (which for the given reactions is expected to approach the non-relativistic exact many-body limit to an RMSD of $\approx$1 kcal/mol\cite{karton2006w4,karton2011w4}---all TPSS results, approximated or not, are $\geq12$ kcal/mol away from this in RMSD).
Curiously, in this particular test even PBE slightly outperforms all the TPSS variants.

As mentioned, almost all practically useful mGGA functionals are parametrized in terms of the $\tau$-intermediate, which is inaccessible to the DF-JX approximation, rather than the $\upsilon$-intermediate.
For this reason, in order to test the practical feasibility of the reported approach, we employed a kinetic energy functional $\tau(\rho,\sigma,\upsilon)$ in order to evaluate an approximate kinetic energy density to use in standard functionals, as suggested by Perdew and Constantin.\cite{perdew2007laplacian}
Concretely, we use LL-TPSS (laplacian-level TPSS), which is the PC07 kinetic energy functional in conjunction with the regular TPSS functional.\cite{perdew2007laplacian}
So far this approach has neither been thoroughly tested, nor found much practical use---probably due to the fact that in the context of a regular density expansion, computing $\upsilon$ is more expensive than computing $\tau$, and offers no advantage over the $\tau$-form of mGGAs we are aware of.

Our hope was that the LL-translation approach via kinetic energy functionals\cite{perdew2007laplacian}  would allow for a straight-forward translation of regular $\tau$-form mGGAs into a $\upsilon$-form which can be employed together with the DF-JX approximation.
Unfortunately, our calculations showed this to not be the case: While the deviations of LL-TPSS and TPSS from CCSD(T) references are similar (similarly large),
the results obtained from the LL-translated TPSS functional and its parent regular $\tau$-mGGA form show significant differences on the order of 4 kcal/mol RMSD in our test. 
This would make blindly trusting LL-translated functionals unadvisable, although in preliminary tests the optimized geometries produced by TPSS and LL-TPSS appeared to be virtually indistinguishable.
In any case, for this reason either extended benchmark calculations to confidently establish the translated functionals' accuracy with respect to high-level references, or, better, direct re-parametrizations of the target functionals in terms of $\upsilon$ rather than $\tau$ would be desirable.
We will attempt to follow both approaches in the future.

\section{Conclusions and Outlook}
While our numerical tests cast some doubts on the reliability of simply \emph{translating} $\tau$-mGGAs into $\upsilon$-mGGAs via kinetic energy functionals\cite{perdew2007laplacian}, the presented $\upsilon$-form mGGA approach in conjunction with the DF-JX approximation may yet turn into a powerful practical tool, due to its high computational efficiency: with this combination, the use of mGGAs is only minimally more expensive than regular GGA functionals (Tab.~\ref{tab:timing}), and as mentioned, we expect it to be entirely feasible to reparametrize potent mGGAs such as MN-15L\cite{truhlar:mn15l} from a $\epsilon(\rho,\sigma,\tau)$ form into a $\epsilon(\rho,\sigma,\upsilon)$-form.
We see particular promise in the computation of transition states, intermediates, and reaction paths in complex reaction networks of small and medium sized molecules, as in such calculations typically several thousand single-point energy+gradient computations are required.
Tasks such as these are the ideal targets for the reported approach, and we will elsewhere report our progress in developing it into an effective method in the arsenal of quantum chemistry.

\section{Acknowledgements}
We acknowledge funding for this project from (a) grant NSF CHE 1263053 for the REU stay of Alyssa V. Bienvenu at the Pennsylvania State University in summer 2016, during which the research of this project was completed; (b) the Roberts Fellowship and a Distinguished Graduate Fellowship for Alyssa V. Bienvenu; and (c) a startup fund for the Gerald Knizia group from the Pennsylvania State University.

\bibliography{refs}


\end{document}